\documentclass[runningheads]{svmult}

\usepackage{makeidx}   % allows index generation
\usepackage{graphicx}  % standard LaTeX graphics tool
\usepackage{subeqnar}  % subnumbers individual equations
\usepackage{multicol}  % used for the two-column index
\usepackage{physprbb}  % modified textarea for proceedings,
\def\bra#1{{\langle #1 |}}
\def\ket#1{{| #1 \rangle}}
\def\bracket#1#2{{\langle #1 | #2 \rangle}}
\def\sx{{\hat\sigma_x}}
\def\sy{{\hat\sigma_y}}
\def\sz{{\hat\sigma_z}}
\def\id{{\hat 1}}
\def\tr{{\rm Tr}}
\def\H{{\hat H}}
\def\U{{\hat U}}
\def\O{{\hat O}}
\def\P{{\hat{\cal P}}}
\def\A{{\hat A}}
\def\Adag{{\hat A^\dagger}}
\def\L{{\hat L}}
\def\Ldag{{\hat L^\dagger}}

\begin{document}
\title*{Decoherence and quantum trajectories}
\toctitle{Decoherence and quantum trajectories}
\author{Todd A. Brun}
\authorrunning{Todd A. Brun}
\institute{Institute for Advanced Study, Princeton, NJ  08540 }

\maketitle

\begin{abstract}
Decoherence is the process by which quantum systems interact and
become correlated with their external environments; quantum
trajectories are a powerful technique by which decohering systems
can be resolved into stochastic evolutions, conditioned on different
possible ``measurements'' of the environment.  By calling on
recently-developed tools from quantum information theory, we can
analyze simplified models of decoherence, explicitly quantifying
the flow of information and randomness between the system, the
environment, and potential observers.
\end{abstract}

\section{Introduction}

In the last twenty years, the concept of {\it decoherence} has gradually
grown to wide acceptance in the description of open quantum systems:
systems which interact with an external environment \cite{Zurek,Joos}.
Such open systems are ubiquitous in nature.  Almost no systems can
be considered to be truly isolated, with the possible exception
of the universe as a whole.  One of the most difficult tasks of
experimenters is to insulate the systems they study from the noisy
effects of the environment, in order to see quantum effects (such as
interference and entanglement) which would otherwise be masked from
us.  This concealing effect of decoherence is the main reason quantum
mechanics was discovered only recently in history:  only microscopic
systems can be isolated sufficiently well to exhibit quantum effects.

Over the last ten years the theory of {\it quantum trajectories}
has been developed by a wide variety of authors
\cite{Carmichael,Dalibard,Dum,Gardiner,Gisin1,Diosi,Gisin2,Schack}
for a variety of purposes,
including the ability to model continuously monitored open systems
\cite{Carmichael,Dum,Gardiner}, improved numerical calculation
\cite{Dalibard,Schack}, and insight into the problem
of quantum measurement \cite{Gisin1,Diosi,Gisin2}.  One of the
most important benefits of quantum trajectories is that they give
a wide range of different descriptions for decoherent systems.

Even more recently, there has been an explosion of interest in
{\it quantum information theory}.  This has been largely stimulated
by interest in quantum computers, and their potential to solve
otherwise intractable problems; but the field has quickly been
seen to give a new paradigm for the study of quantum systems,
by abstracting their quantum properties from the details of their
physical embodiments.

An obvious possibility then suggests itself:  to use these new tools
of quantum information theory to analyze open quantum systems,
giving insights into the nature of decoherence and quantum trajectories.
First, let us review some simple ideas from quantum information,
which will suffice to construct simple models of systems and
environments.  With these models, we can explicitly track the flow
of information and randomness in quantum open systems.
(For a good general source on quantum computation and
information, see \cite{NielsenChuang} and references therein;
for a more complete application to quantum trajectories,
see \cite{Brun1}.)

\subsection{Q-bits and gates}

The simplest possible quantum mechanical system is a two-level atom or
{\it q-bit}, which has a two-dimensional Hilbert space ${\cal H}_2$.
There are many physical embodiments of such a system:  the spin of a
spin-1/2 particle, the polarization states of a photon, two hyperfine
states of a trapped atom or ion, two neighboring levels of a Rydberg atom,
the presence or absence of a photon in a microcavity, etc.  For our
purposes, the particular physical embodiment is irrelevant.  Q-bits
were introduced in quantum information theory by analogy with
{\it classical} bits, which can take two values, $0$ or $1$.  Just
as a q-bit is the simplest imaginable quantum system, a classical bit
(or c-bit) is the simplest system which can contain any information.

By convention, we choose a particular basis and label its basis states
$\ket0$ and $\ket1$, which we define to be the eigenstates of the Pauli
spin matrix $\sz$ with eigenvalues $+1$ and $-1$, respectively.
We similarly define the other Pauli operators $\sx,\sy$; linear combinations
of these, together with the identity $\id$, are sufficient to produce any
operator on a single q-bit.

The most general pure state of a q-bit is
\begin{equation}
\ket\psi = \alpha\ket0 + \beta\ket1,\ \ |\alpha|^2 + |\beta|^2 = 1 \;.
\label{qbit_pure}
\end{equation}
A global phase may be assigned arbitrarily, so all physically distinct
pure states of a single q-bit form a two-parameter space.

If we allow states to be {\it mixed}, we represent a q-bit by a density
matrix $\rho$; the most general density matrix can be written
\begin{equation}
\rho = p \ket\psi\bra\psi + (1-p) \ket{\bar\psi}\bra{\bar\psi} \;,
\label{orthog_rho}
\end{equation}
where $\ket\psi$ and $\ket{\bar\psi}$ are two orthogonal pure states,
$\bracket\psi{\bar\psi}=0$.  The mixed states of a q-bit
form a three parameter family.

For {\it two} q-bits, the Hilbert space ${\cal H}_2\otimes{\cal H}_2$
has a tensor-product basis
\begin{equation}
\ket{i}_A\otimes\ket{j}_B \equiv \ket{ij}_{AB} \;,
  \ \ i,j \in \{0,1\} \;;
\end{equation}
similarly, for $N$ q-bits we can define a basis
$\{ \ket{i_{N-1}i_{N-2}\cdots i_0} \}$, $i_k=0,1$.

All states evolve according to the Schr\"odinger equation with some
Hamiltonian $\H(t)$,
\begin{equation}
{d\ket\psi\over dt} = - {i\over\hbar} \H(t) \ket\psi.
\label{schrodinger}
\end{equation}
(Henceforth, I will assume $\hbar=1$.)
Over a finite time this is equivalent to applying a unitary operator
$\U$ to the state $\ket\psi$,
\begin{equation}
\U = {\rm T:} \exp \left\{ - i \int_{t_0}^{t_f} dt\, \H(t)
  \right\} : \;,
\end{equation}
where ${\rm T:}:$ indicates that the integral should be taken in a
time-ordered sense, with early to late times being composed from right
to left.  For the models I consider in this paper I will treat all time
evolution at the level of unitary transformations rather than explicitly
solving the Schr\"odinger equation, so time can be treated as a discrete
variable
\begin{equation}
\ket{\psi_n} = \U_n \U_{n-1} \cdots \U_1 \ket{\psi_0} \;.
\end{equation}
For a mixed state $\rho$, Schr\"odinger time evolution is
equivalent to $\rho \rightarrow \U \rho \U^\dagger$.

Because these unitary transformations are discrete, a transformation
on only one or a few q-bits is analogous to a {\it logic gate} in
classical information theory.  Typical classical gates are the NOT
(which affects only a single bit), and the AND and the OR (which affect
two).  Such gates are defined by a truth table, which gives the output
for given values of the input bits.

In the quantum case, there is a continuum of possible unitary
transformations.  I will consider only a
limited set of two-bit transformations in this paper, and no
transformations involving more than two q-bits; but these models
are readily generalized to more complex situations.  Let us examine
a couple of examples of quantum two-bit transformations.  The
controlled-NOT gate (or CNOT) is widely used in quantum computation;
it can be defined by its action on the tensor-product basis vectors:
\begin{equation}
\U_{\rm CNOT} \ket{ij} = \ket{i(i\oplus j)} \;,
\label{cnot_def}
\end{equation}
where $\oplus$ denotes addition modulo 2.
If the first bit is in state $\ket0$ this gate leaves the second bit
unchanged; if the first bit is in state $\ket1$ the second bit is
flipped $\ket0 \leftrightarrow \ket1$.  Hence the name:  whether a NOT
gate is performed on the second bit is {\it controlled} by the first bit.

Another important gate in quantum computation is the SWAP; applied
to the tensor-product basis vectors it gives
\begin{equation}
\U_{\rm SWAP} \ket{ij} = \ket{ji} \;.
\label{swap_def}
\end{equation}
As the name suggests, the SWAP gate just exchanges the states of the
two bits:  $\U_{\rm SWAP}(\ket\psi\otimes\ket\phi)
= \ket\phi\otimes\ket\psi$.

CNOT and SWAP are examples of two-bit {\it quantum gates}.  Such gates are
of tremendous importance in the theory of quantum computation.  More
general unitary transformations can be built up by applying a succession
of such quantum gates to the q-bits which make up the system.  Such
a succession of quantum gates is called a {\it quantum circuit}.

\subsection{Projective measurements}

In the standard description of quantum mechanics, observables are identified
with Hermitian operators $\O=\O^\dagger$.  A measurement of a system initially
in state $\ket\psi$ returns an eigenvalue
$o_n$ of $\O$, and leaves the system in the eigenstate
$\P_n\ket\psi/\sqrt{p_n}$, where 
$\P_n$ is the projector onto the eigenspace corresponding to eigenvalue
$o_n$; the probability of the measurement outcome is
$p_n = \bra\psi\P_n\ket\psi$.
For a mixed state $\rho$ the probability of
outcome $n$ is $p_n = \tr\{\P_n\rho\}$ and the state after the measurement
is $\P_n\rho\P_n/p_n$.

Because two observables with the same eigenspaces are completely equivalent
to each other (as far as measurement probabilities and outcomes
are concerned), we will not worry about the exact choice
of Hermitian operator $\O$; instead, we will
choose a complete set of orthogonal projections $\{\P_n\}$ which
represent the possible measurement outcomes.  These satisfy
\begin{equation}
\P_n \P_{n'} = \P_n \delta_{nn'} \;, \ \ \sum_n \P_n = \id \;.
\label{orthogonal_decomp}
\end{equation}
A set of projection operators which obey (\ref{orthogonal_decomp}) is
often referred to as an {\it orthogonal decomposition of the identity}.  For
a single q-bit, the only nontrivial measurements have exactly two outcomes,
which we label $+$ and $-$, with probabilities $p_+$ and $p_-$ and
associated projectors of the form
\begin{equation}
\P_\pm = \frac{ \id \pm {\vec n}\cdot{\hat{\vec\sigma}} }{2}
  = \ket{\psi_\pm}\bra{\psi_\pm} \;,
\end{equation}
where $\vec n$ is a unit 3-vector and ${\hat{\vec\sigma}}=(\sx,\sy,\sz)$.
The two projectors sum to the identity operator, $\P_++\P_-=\id$.
The average information obtained from a projective measurement on a q-bit is
the {\it Shannon entropy} for the two measurement outcomes:
\begin{equation}
S_{\rm meas} = - p_+ \log_2 p_+ - p_- \log_2 p_- \;.
\label{shannon}
\end{equation}
The maximum information gain is precisely one bit, when $p_+=p_-=1/2$,
and the minimum is zero bits when either $p_+$ or $p_-$ is 0.
After the measurement, the state is left in an eigenstate of
$\P_+$ or $\P_-$, so repeating the measurement will result in the
same outcome.  This repeatability is one of the most important features
of projective measurements.

\section{Quantifying quantum information}

\subsection{How much information in a q-bit?}

The Shannon entropy is the standard measure used in ordinary
classical information theory to quantify the information gain from a
random source.  It is interesting to see how far we can get in
quantifying {\it quantum} information, using only the tools that I've
described so far.

To begin with, let's ask a question that's been around from the beginning
of quantum information theory:  {\it how much information
is contained in a quantum state?}  We can consider two possible answers
to this question.  In the first place, we could measure the quantum
system in question.  The information gained is quantified by the Shannon
entropy of the outcome.  Suppose that our system is a q-bit.  Then the
measurement has at most two possible outcomes, for a maximum information
gain of one bit.

On the other hand, there are an infinite number of possible states
(\ref{qbit_pure}) for a q-bit, forming a continuum of states parametrized
by the complex numbers $\alpha$ and $\beta$.  To completely specify
$\alpha$ and $\beta$ (for instance if we wanted to prepare the system
in a particular state) would take an {\it infinite} number of bits.  Thus,
it seems that a quantum system can contain far more information than it
is possible to extract.

This seeming paradox can be resolved by distinguishing between the
physical system and the state, which is a {\it description} of
that system.  It is not necessarily surprising that it might take far
more information to give a complete description of a system than it is
possible to extract from that system.  For instance, consider a classical
bit $x$, which can take the values $0$ or $1$.  This bit could be chosen
randomly according to a probability distribution $p(x)$, which requires
specifying the values $p(0)$ or $p(1)$.  Since these are real
numbers, to describe them in this case too would require an infinite
number of bits.

It is good to bear this distinction between system and state in mind,
since it is not always completely clear in quantum information whether
one is manipulating the system or the state.  For instance, in the
well-known protocol of quantum teleportation, it is not the physical
system which is transfered, but rather its state.

\subsection{Shannon and von Neumann entropy}

The Shannon entropy, or information gain, depends strongly on both the
state and the choice of measurement.  For a q-bit, this ranges from
0 bits (representing a determined outcome) and 1 bit (representing a
maximally uncertain outcome).  Because the probabilities depend on the
choice of measurement, we cannot associate a definite value of the
Shannon entropy with the state.  This is unlike the case of a classical
probability distribution, where the Shannon entropy is unique.

We can, however, ask the minimum and maximum values of the Shannon
entropy for a given state.  We consider all possible measurements
which are {\it maximally fine-grained}, i.e., which have $D$ distinct
outcomes for a $D$-dimensional system -- two, in the case of a q-bit.
For any state of a system with a Hilbert space of dimension $D$, the
{\it maximum} Shannon entropy is $\log_2 D$.  That means that for any
state, we can find a measurement which is maximally uncertain.

The minimum, however, is quite different.  For a pure state, it is always
possible to find a measurement with Shannon entropy 0.  For a mixed state
this is not true.  For any mixed state $\rho$, the minimum Shannon
entropy of a fine-grained measurement is greater than 0.

What is the interpretation of this minimum entropy?  We associate it
with our {\it ignorance} of a system.  For a pure state, this minimum
entropy is zero, which we take to mean that we know {\it as much as possible}
about this system -- we have {\it maximal knowledge}, or {\it minimal
ignorance}.  For a mixed state, however, our ignorance is not minimal --
we could learn more.

This minimum value of the Shannon entropy has a
relatively simple formula:
\begin{equation}
S(\rho) = - \tr\{ \rho\log_2 \rho \} \;.
\end{equation}
This is the {\it von Neumann entropy}.  It vanishes for pure
states, and takes a maximum value of $\log_2 D$ for the
{\it maximally mixed} state $\id/D$.

\subsection{Randomness}

Unlike the classical case, having maximal information about
a system does {\it not} imply that we can predict the outcome of any
measurement.  It means only that there is {\it some} measurement which we
can predict with certainty.  For most measurements, the Shannon entropy
will not vanish.

Suppose we have a q-bit in the state $\alpha\ket0 + \beta\ket1$, and
we carry out a measurement in the $\ket0,\ket1$ basis.  The Shannon
entropy for this measurement is $-|\alpha|^2 \log_2 |\alpha|^2
- |\beta|^2 \log_2 |\beta|^2$, which will vanish only if either $\alpha$
or $\beta$ is zero.  I described this before as the information gained
from the measurement.

One might logically ask at this point:  information about what?  The
q-bit was initially in a pure state, which I have just stated to represent
maximum knowledge.  After the measurement, the q-bit is still in a pure
state.  I cannot, therefore, have gained any further information
about the system.  We can only conclude that the information
I acquired by carrying out the measurement represents pure {\it randomness}.

We can illustrate this rather spectacularly by considering a q-bit
initially in the state $(\ket0+\ket1)/\sqrt2$.  We alternately
measure the system in the bases $\ket0,\ket1$ and
$(\ket0\pm\ket1)/\sqrt2$.  Each time we measure the system, the outcome
is completely indeterminate; we gain exactly one bit of information.
By continuing this procedure as long as we like, we can gain as many
bits of information as we wish.  But none of these bits actually
represent information about the system.  The system starts, and remains
for all time, in a pure state.  All of the bits we acquire are pure
randomness, and the q-bit and measurements form a {\it randomness pump}.

\section{A Simple Plan}

Let us now consider a simple model of a quantum process, which forms
the basis of the rest of this talk.  Consider a very simple quantum
system:  a single q-bit, which begins in a pure state $\ket{\psi_0}$.
We then send in a second q-bit, the {\it probe}, in state $\ket{\phi_0}$.
The two q-bits interact briefly, for a period $\delta t$, before flying
apart again; this interaction causes them to undergo a joint unitary
transformation $\U$.  After they have interacted, we may intercept the
probe and measure it, with projection operators $\P_+$ and $\P_-$
representing the two possible outcomes of the measurement.  A schematic
picture of this process is given in figure 1.

\begin{figure}[b]
\begin{center}
\includegraphics[width=.6\textwidth]{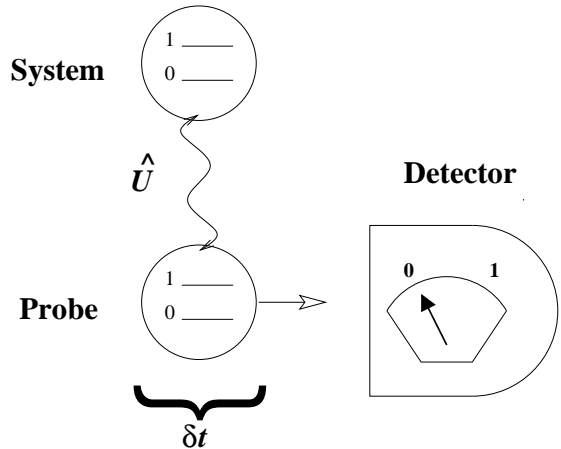}
\end{center}
\caption[]{This is a schematic diagram of the type of model used in
this paper.  The system is a single two-level system, or q-bit.  It
interacts briefly with a probe q-bit over a time interval
$\delta t$, and the probe bit is subsequently measured}
\label{fig1}
\end{figure}

The initial state of the system and probe can be written as a simple
tensor product $\ket{\psi_0}_s \otimes \ket{\phi_0}_p$.  After the
system and probe have interacted, however, the new joint state $\ket\Psi$
will generally no longer be a product.
A state of this type, which cannot be written as a product, is said
to be {\it entangled}.  Such states have many curious properties, without
classical analogues.

Even though the joint state $\ket{\Psi}$ is pure, we cannot
describe either the system or the probe alone by a pure state.  It {\it is}
possible, however, to describe them by a {\it mixed} state, by finding
the {\it reduced density operator}.  The reduced density operator $\rho_s$
for the system is found by taking a partial trace of the joint state
over the probe degree of freedom:
\begin{equation}
\rho_s = \tr_p \{ \ket\Psi \bra\Psi \} \;.
\end{equation}
This mixed state $\rho_s$ gives exactly the same predictions as the joint
state $\ket\Psi$ for any measurement which is restricted to the system
alone.  We can, of course, find a similar reduced density matrix $\rho_p$
for the probe by taking a partial trace over the system.

Provided that the joint state $\ket\Psi$ is pure, the mixed states
$\rho_s$ and $\rho_p$ must have the same von Neumann entropy:
$S(\rho_s) = S(\rho_p)$.  (This is true in general, not just for q-bits.)
Because this quantity is the same whichever subsystem we trace out,
and because it vanishes for product states, it is widely used as a measure
of entanglement for pure states:  the {\it entropy of entanglement},
$S_E(\ket\Psi)$.

Suppose now that our system and probe have interacted and are in an
entangled state $\ket{\Psi}$.  What happens if we measure the probe?
As mentioned above, the measurement is represented by the two projection
operators $\P_{\pm}$ which sum to the identity $\P_++\P_-=\id$.
Because these are projectors onto a two-dimensional Hilbert space,
each of them projects onto a one-dimensional subspace.  We can therefore
write them as $\P_+ = \ket+\bra+$, $\P_- = \ket-\bra-$.  After allowing
the system and probe to interact, and then measuring the probe, the
system and probe will be in one of two possible joint states:
\begin{eqnarray}
(\id \otimes \P_\pm) \ket\Psi
  &=& (\id \otimes \P_\pm) \U \ket{\psi_0} \otimes \ket{\phi_0} \nonumber\\
  &=& \A_\pm \ket{\psi_0} \otimes \ket{\pm} \;,
\end{eqnarray}
(where we have not renormalized the final state).  The system and probe are
once more in a product state.  The operators $\A_\pm$
are determined by the unitary transformation $\U$ and the initial state
of the probe $\ket{\phi_0}$.  The probabilities of the two outcomes are
\begin{eqnarray}
p_\pm &=& (\bra{\psi_0}\otimes\bra{\phi_0}) \U^\dagger (\id \otimes \P_\pm)
  \U (\ket{\psi_0} \otimes \ket{\phi_0}) \nonumber\\
&=& \bra{\psi_0} \Adag_\pm \A_\pm \ket{\psi_0} \;.
\end{eqnarray}
The fact that these two probabilities must add to 1 for any state
$\ket{\psi_0}$ implies that $\Adag_+\A_+ + \Adag_-\A_- = \id$.

If we discard the probe after the measurement and renormalize the state,
the system is left in the new state $\A_\pm\ket{\psi_0}/\sqrt{p_\pm}$.
This is quite similar to the effects of a projective measurement;
indeed, if $\A_\pm$ are projectors, this reduces to the usual formula
for a projective measurement.  Because of this, and because we are
indirectly acquiring information about the system by measuring the
probe, this is commonly referred to as a {\it generalized measurement}.

How much information is gained in such a generalized measurement?  We
can calculate this in exactly the same way as for a projective measurement.
The Shannon entropy of the generalized measurement is
$- p_+ \log_2 p_+ - p_- \log_2 p_-$.  This must obviously be greater
than the entropy of entanglement $S_E(\ket\Psi)$.  We can choose
projectors $\P_\pm$ to {\it minimize} the Shannon entropy by writing
the state $\ket\Psi$ in {\it Schmidt form}:
\begin{equation}
\ket\Psi = \sqrt{p_+} \ket{+}_s \otimes \ket{+}_p
  + \sqrt{p_-} \ket{-}_s \otimes \ket{-}_p \;.
\end{equation}
Choosing the right Schmidt bases $\ket{\pm}_{s,p}$ requires us to
know the initial states $\ket{\psi_0}$ and $\ket{\phi_0}$ and
the unitary transformation $\U$.

Of course, in the case described above, the system starts and ends in
a pure state; so this generalized measurement also generates randomness.
However, it is certainly capable of giving information about the
system.  Suppose that the initial state of the system is maximally mixed:
$\rho_s = \id/2$, with $S(\rho_s)=1$.  If we have it interact
with the probe and carry out the measurement, then with probabilities
\begin{equation}
p_\pm = \tr\{ \A_\pm \rho_s \Adag_\pm \}
\end{equation}
the system will be left in one of the states
\begin{equation}
\rho_\pm = \A_\pm \rho_s \Adag_\pm/p_\pm
  = \A_\pm\Adag_\pm/2p_\pm \;.
\end{equation}
In general, neither of these states will be maximally mixed.  The
entropy of the system will be diminished by an average amount
\begin{equation}
\Delta S = 1 - p_+ S(\rho_+) - p_- S(\rho_-) \;.
\end{equation}
This represents the actual information gained about the system.  This
must always be less than or equal to the Shannon entropy of the
generalized measurement, $- p_+ \log_2 p_+ - p_- \log_2 p_-$.

Let's look at a couple of examples to see how this works.  Suppose the
system is initially in the maximally mixed state $\rho_s = \id/2$;
the probe is in the pure state $\ket{\phi_0}=\ket0$; and we measure
the probe using projectors onto the states $\{\ket0,\ket1\}$.  We
let the interaction $\U$ be the CNOT gate from (\ref{cnot_def}).
In this case, we gain exactly 1 bit of information about the system,
equal to the Shannon entropy of the measurement, and leaving the
system in a pure state $\ket0$ or $\ket1$.

Suppose that we keep the same initial states and interaction, but instead
make the measurement given by $\P_\pm = \ket\pm \bra\pm$ where
$\ket\pm = (\ket0 \pm \ket1)/\sqrt2$.  In this case, the Shannon entropy
is still 1 bit, but we now gain no information about the system; it is
left in exactly the same state as it started, the maximally mixed state
$\rho_s = \id/2$.

If we further generalize this scheme and allow the initial state of
the probe to be mixed, then it is actually possible to {\it lose}
information about the system; $\Delta S$ can be negative.  For instance,
an initial pure state for the system can become mixed, due to noise
from interacting with a mixed environment.

With this very simple model of a quantum process, we can build up everything
we need to understand both decoherence and quantum trajectories.  We
examine them both in the next two sections.

\section{Decoherence}

In discussing quantum evolution it is usually assumed that the quantum
system is very well isolated from the rest of the world.  This is a
useful idealization, but it is rarely realized in practice, even
in the laboratory.  In fact, most systems interact at least
weakly with external degrees of freedom \cite{Zurek,Joos}.
This is the process of {\it decoherence}.

One way of taking this into account is to include
a model of these external degrees
of freedom in our description.  Let us assume that in addition
to the system in state $\ket\psi \in {\cal H}_S$
there is an external {\it environment}
in state $\ket{E}\in{\cal H}_E$.

Systems which interact with their environments are said to be {\it open}.
Most real physical environments are extremely complicated, and the
interactions between systems and environments are often poorly understood.
In analyzing open systems, one often makes the approximation of assuming
a simple, analytically solvable form for the environment degrees of
freedom.

For this paper, I will assume that both the system and
the environment consist solely of q-bits.  I will also assume a simple
form of interaction, namely that the system q-bit interacts with one
environment q-bit at a time, and that after interacting they never come
into contact again; and that the environment q-bits have no Hamiltonian
of their own. This may seem
ridiculously oversimplified, but in fact it suffices to demonstrate most
of the physics exhibited by much more realistic descriptions.

For this type of model, the Hilbert space of the system is
${\cal H}_2$ and the Hilbert space of the environment
is ${\cal H}_E = {\cal H}_2 \otimes {\cal H}_2 \otimes \cdots$.
I will assume that all the environment q-bits start in some
pure initial state $\ket{\phi_0}$, usually $\ket0$, though further
elaborations of the model could include
other pure-state and mixed-state environments.  These environment
q-bits play a role exactly like the probe in section 3,
except that they are never measured.

We can describe this as an effective process on the system alone.
After each interaction, the system's reduced state $\rho$
undergoes the evolution
\begin{equation}
\rho \rightarrow \sum_{k=\pm} \A_k \rho \Adag_k \;,
\label{decoherence}
\end{equation}
so that after $t$ steps the state becomes
\begin{equation}
\rho(t) = \sum_{k_1,\ldots,k_t=\pm} \A_{k_t} \cdots \A_{k_1}
  \rho(0) \Adag_{k_1} \cdots \Adag_{k_t} \;.
\end{equation}
The operators $\A_\pm$ depend on the choice of $\U$ and $\ket{\phi_0}$;
they are determined just as in the generalized measurement case described
in section 3.  However, this determination is not unique;
many choices of $\A_k$ yield the same process (\ref{decoherence}).

An interesting case is when the interaction $\U$ is {\it weak},
that is, close to the identity:
\begin{equation}
\U = \exp(-i\epsilon\H) \approx \id - i \epsilon\H - (\epsilon\H)^2/2 \;,
\end{equation}
where $|\H| \sim O(1), |\epsilon| \ll 1, \H = \H^\dagger$.
If the environment q-bits arrive with an average separation of $\delta t$,
and $\epsilon$ is sufficiently small, then the system state $\rho$
will approximately obey a continuous evolution equation
\begin{equation}
{d\rho\over dt} = - i[\H_{\rm eff},\rho] +
  \L \rho \Ldag - (1/2) \Ldag\L\rho
  - (1/2)\rho\Ldag\L  \;.
\label{Lindblad}
\end{equation}
This is a {\it Lindblad master equation} \cite{Lindblad}.
In terms of $\H$, $\H_{\rm eff} \propto \bra{\phi_0}\H\ket{\phi_0}$,
and $\L \propto \bra{\bar{\phi_0}}\H\ket{\phi_0}$.

Let's consider a concrete example.  Suppose the interaction is
$\U=\exp(-i\epsilon\H)$ with $\H = \sz \otimes \sx$, and the environment
bits are initially in state $\ket0$.  Then the reduced density matrix
for the system alone will obey (\ref{Lindblad}) with $\H_{\rm eff} = 0$
and $\L = (\epsilon/\sqrt{\delta t})\sz$.  This master equation will
cause a system initially in the pure state $\alpha\ket0 + \beta\ket1$
to evolve in the long time limit to the mixed state
$\rho = |\alpha|^2 \ket0\bra0 + |\beta|^2 \ket1\bra1$.  As the system
state becomes mixed, its von~Neumann entropy grows, reflecting a
gradual loss of information about the system, or (alternatively) a
growing entanglement of the system with the environment.  Different
interactions or environment states of course will lead to different
master equations.

\section{Quantum trajectories}

We can readily generalize our model of decoherence by supposing that
we have experimental access to the q-bits of the environment.  After
each bit has interacted with the system, we measure it using some
predefined projective measurement; based on the outcome of this measurement,
we update our knowledge of the system state.  The series of decohering
interactions then becomes a series of generalized measurements, as
described in section 3.

The evolution of the system state is no longer deterministic; it becomes
stochastic due to the randomness of the measurement
outcomes.  We also now acquire information about the system as it is lost
to the environment.  Given perfect measurements of the environment, the
system will remain always in a pure state.

Rather than the evolution (\ref{decoherence}), the system now undergoes
\begin{equation}
\ket{\psi} \rightarrow \A_\pm \ket{\psi}/\sqrt{p_\pm}
\end{equation}
with probabilities $p_\pm = \bra{\psi}\Adag_\pm\A_\pm\ket{\psi}$.
After $t$ steps, the state will have become
\begin{equation}
\ket{\psi(t)} = \A_{k_t}\cdots\A_{k_1}\ket{\psi(0)}/
  \sqrt{p_{k_1,\ldots,k_t}} \;,
\end{equation}
where the $k_i$ take the values $\pm$.
Note that, while there was an ambiguity in $\A_\pm$ for
our decoherence model, by choosing a particular measurement we fix
a particular choice of $\A_\pm$.

This evolution becomes more interesting when the interaction is {\it weak},
as described in the previous section.  In this case, our series of
generalized measurements are {\it weak measurements}
\cite{Vaidman} -- on average, they
disturb the system state very little, but also give very little
information.  The effective evolution of the system becomes
approximately continuous, but rather than a master equation, it is
a {\it stochastic Schr\"odinger equation}.

Let us consider the same system described at the end of section 4,
but now including a measurement of each environment q-bit after it has
interacted with the system.  After the interaction, the joint state of
the system and environment bit is
\begin{equation}
\U \ket\psi \otimes \ket0 \approx (1-\epsilon^2/2) \ket\psi \otimes \ket0
  - i \epsilon \sz \ket\psi \otimes \ket1 \;.
\end{equation}
Suppose that we measure the environment bit in the $\{\ket0,\ket1\}$
basis.  Then with probability $p_0 \approx 1-\epsilon^2$ the bit
will be found in state $\ket0$ and the system state will remain
unchanged.  With a small probability $p_1 \approx \epsilon^2$, however,
the environment bit will be found in state $\ket1$ and the state of
the system will change to $\sz\ket\psi$.  This type of evolution can
be approximated as a {\it quantum jump equation}:
\begin{equation}
\ket{d\psi} = (\sz - \id)\ket\psi dN \;,
\end{equation}
where $dN$ is a stochastic differential variable which is 0 most
of the time, but occasionally (with probability of $\epsilon^2$ in
each interval $\delta t$) becomes 1.  Writing this in terms of the
statistical mean $M[\cdot]$,
\begin{equation}
dN^2 = dN \;,\ \ \ M[dN] = (\epsilon^2/\delta t) dt \;.
\end{equation}

Instead of the measurement above, we might instead measure the environment
bits in the basis $\ket\pm = (\ket0 \pm \ket1)/\sqrt2$.  In terms of
this basis, the joint state becomes
\begin{equation}
\U \ket\psi \otimes \ket0
  = {1\over\sqrt2} \exp(-i\epsilon\sz)\ket\psi \otimes \ket+
  + {1\over\sqrt2} \exp(+i\epsilon\sz)\ket\psi \otimes \ket- \;.
\end{equation}
The system state undergoes one of two weak unitary transformations,
based on the outcome of the measurement; the two outcomes have
equal probability.  This evolution is approximated by the continuous
{\it quantum state diffusion} equation
\begin{equation}
\ket{d\psi} =  - i \L \ket\psi dW \;,
\end{equation}
where $\L=(\epsilon/\sqrt{\delta t})\sz$ and
$dW$ is a real differential stochastic variable obeying
\begin{equation}
M[dW] = 0 \;, M[dW^2] = dt \;.
\end{equation}

We see how exactly the same physical system can exhibit two very
different-looking evolutions based on the choice of environmental
measurement.  In both cases, if we average over all possible trajectories
we recover the solution of the master equation (\ref{Lindblad}):
\begin{equation}
M[\ket{\psi(t)}\bra{\psi(t)}] = \rho(t) \;.
\end{equation}
Because of this, these different stochastic Schr\"odinger equations
are often referred to as different {\it unravelings} of the
master equation.

\section{Quantum trajectories and decoherent histories}

As is clear from the previous sections, the formalism of quantum
trajectories calls on nothing more than standard quantum mechanics,
and as framed above is in no way an alternative theory or interpretation.
Everything can be described solely in terms of measurements and unitary
transformations, the building blocks of the usual Copenhagen interpretation.

However, many people have expressed dissatisfaction with the
standard interpretation over the years, usually due to
the role of measurement as a fundamental
building block of the theory.  Measuring devices are large, complicated
things, very far from elementary objects; what exactly constitutes a
measurement is never defined; and the use of classical mechanics to
describe the states of measurement devices is not justified.  Presumably
the individual atoms, electrons, photons, etc., which make up a detector
can themselves be described by quantum mechanics.  If this is carried to
its logical conclusion, however, and a Schr\"odinger equation is
constructed for the measurement process, one obtains not classical
behavior, but rather giant macroscopic superpositions such as the famous
Schr\"odinger's cat paradox \cite{Schroedinger}.

One approach to this problem is to retain the usual quantum theory, but
to eliminate measurement as a fundamental concept, finding some other
interpretation for the predicted probabilities.  While many interpretations
follow this approach, the one that is most closely tied to quantum
trajectories is the {\it decoherent} (or {\it consistent}) {\it histories
formalism} of Griffiths, Omn\'es, Gell-Mann and Hartle
\cite{Griffiths,Omnes,GellMann}.  In this formalism,
probabilities are assigned to {\it histories} of events rather than
measurement outcomes at a single time.  These can be
grouped into sets of mutually exclusive, exhaustive histories whose
probabilities sum to 1.  However, not all histories can be assigned
probabilities under this interpretation; only histories which lie in sets
which are {\it consistent}, that is, whose histories do not
exhibit interference with each other, and hence obey the usual
classical probability sum rules.

Each set is basically a choice of description for the quantum system.
For the models considered in this paper, the quantum trajectories correspond
to histories in such a consistent set.  The probabilities of the
histories in the set exactly equal the probabilities of the measurement
outcomes corresponding to a given trajectory.  This equivalence has been
shown between quantum trajectories and consistent sets for certain more
realistic systems, as well \cite{DGHP,Brun2,Brun3}.

For a given quantum system, there can be multiple consistent descriptions
which are {\it incompatible} with each other; that is, unlike in classical
physics, these descriptions cannot be combined into a single, more
finely-grained description.  In quantum trajectories, different unravelings
of the same evolution correspond to such incompatible descriptions.
In both cases, this is an example of the complementarity of quantum
mechanics.

We see, then, that while quantum trajectories can be straightforwardly
defined in terms of standard quantum theory when the environment is
subjected to repeated measurements, even in the absence of such measurements
there is an interpretation of the trajectories in terms of decoherent
histories.  Because the consistency conditions guarantee that the
probability sum rules are obeyed, one can therefore use quantum trajectories
as a calculational tool even in cases where no actual measurements
take place.

\section{Conclusions}

In this paper I have presented a simple model of a system and environment
consisting solely of quantum bits, using no more than single-bit
measurements and two-bit unitary transformations.  The
simplicity of this model makes it particularly suitable for demonstrating
the properties of decoherence and quantum trajectories.
We can quantify the transfer of information from
system to environment, the amount of entanglement, and the randomness
produced by particular choices of measurement.

Quantum trajectories can often simplify the description of an open
quantum system in terms of a stochastically evolving pure state rather
than a density matrix.  While for the q-bit models of this paper there
is no great advantage in doing so, for more complicated systems this
can often make a tremendous practical difference \cite{Schack}.

The ideas behind decoherence and quantum trajectories developed largely
separately from the ideas which have led to the recent explosion of
interest in quantum information; but I would argue that both areas
can contribute much to the understanding of the other.  I hope that
this paper has given support to this view.

\section*{Acknowledgments}

I would like to thank Steve Adler, Howard Carmichael,
Lajos Di\'osi, Bob Griffiths, Jim Hartle, R\"udiger Schack,
Artur Scherer and Andrei Soklakov for their comments, feedback, and
criticisms of the ideas behind this paper; and Hans-Thomas Elze
for his kind invitation to speak at the DICE conference in Piombino,
and for including this work in the resulting volume of lectures.
This work was supported in part by NSF Grant No.~PHY-9900755, by
DOE Grant No.~DE-FG02-90ER40542, and by the Martin A. and Helen Chooljian
Membership in Natural Science, IAS.

\end{document}